# THE WAVE METHOD OF BUILDING COLOR PALETTE AND ITS APPLICATION IN COMPUTER GRAPHICS

Sabo I.I., Lagoda H.R.

Zaporizhzhya National University, Ukraine

## ABSTRACT

This article describes a method of getting a harmonious combination of colors, developed by us on the basis of the relationship of color and acoustic waves. Presents a parallel between harmoniously matched colors and the concept of harmony in music theory (consonance). Describes the physical assumption of the essence of the phenomenon of harmony (consonance). The article also provides algorithm of implementation wave method for the sRGB color model.

**Keywords:** wave method, color palette, color, wave, acoustics, music theory, computer graphics, sRGB.

## INTRODUCTION

In colouristics there are several methods for constructing a color palette based on the arrangement of colors relative to each other in the color circle [1, p. 62]. Harmonious perception of which is not sufficiently substantiated from the physical point of view.

In music theory there is the concept of consonant intervals. Consonances are called intervals, sounding more softly, the sounds of which seem to merge with each other. There are three groups of consonances: very perfect (perfect unison, octave), perfect (perfect quint, perfect quartet) and imperfect (big third, small third, sixth) [2, p. 69]. There is also the concept of a consonant chord - a major or minor triad which consists only consonant intervals [2, p. 85].

Acoustically, the essence of the difference between consonance and dissonance is expressed in different lengths of periods of regularly repeated groups of oscillations [3, p. 18]. The criterion of the difference between consonance and dissonance is the simplicity or complexity of the relationship: simpler relationship is more consonant, more difficult is more dissonant. Numerical proportions can be expressed in two ways: through the ratios of the lengths of the strings or through the ratios of the vibration numbers [3, p. 16 - 17]. In other words, the degree of consonance of two notes is determined by the number of coincidences of the periods of the corresponding harmonic functions of the dependence of sound pressure on time per unit time.

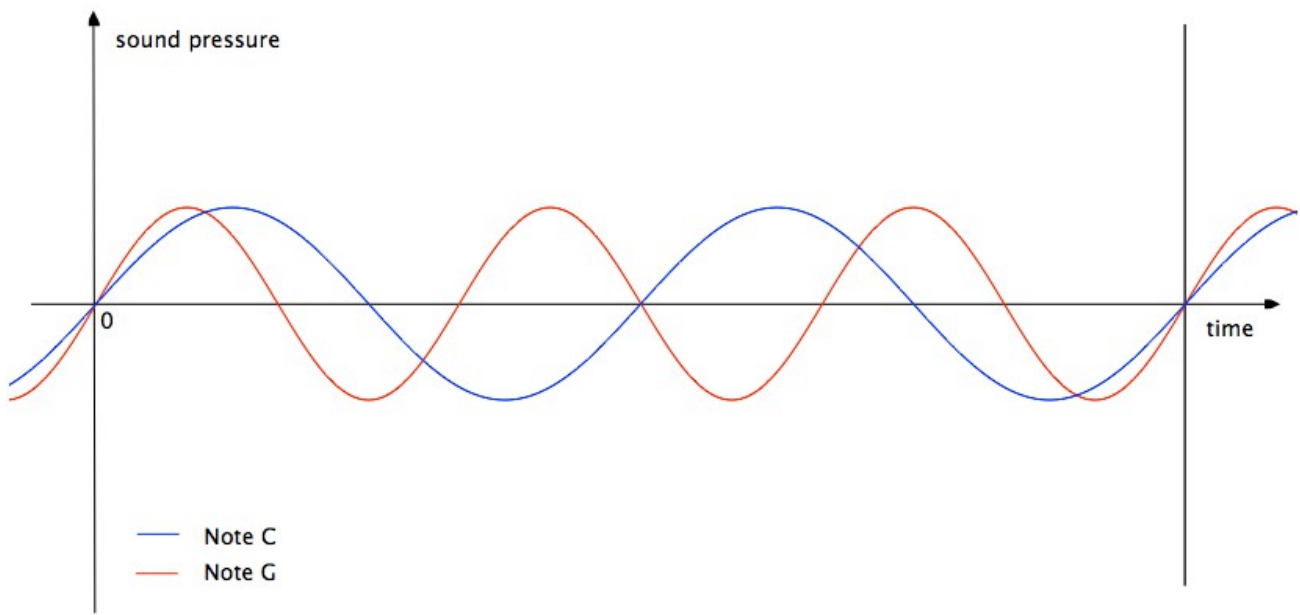

Figure 1: Graph of sound pressure versus time in fixed distance from sound source.

For example, notes C and G (perfect fifth) have the lengths of sound waves differing by 1.5 times. The graphs of the functions of the dependence of the sound pressure of the notes on time intersect on the abscissa axis (sound pressure equal to zero) when the sound pressure function of the note C makes two oscillations, and the function of the note G is three (fig. 1). On figure 1 this moment is marked with a vertical line.

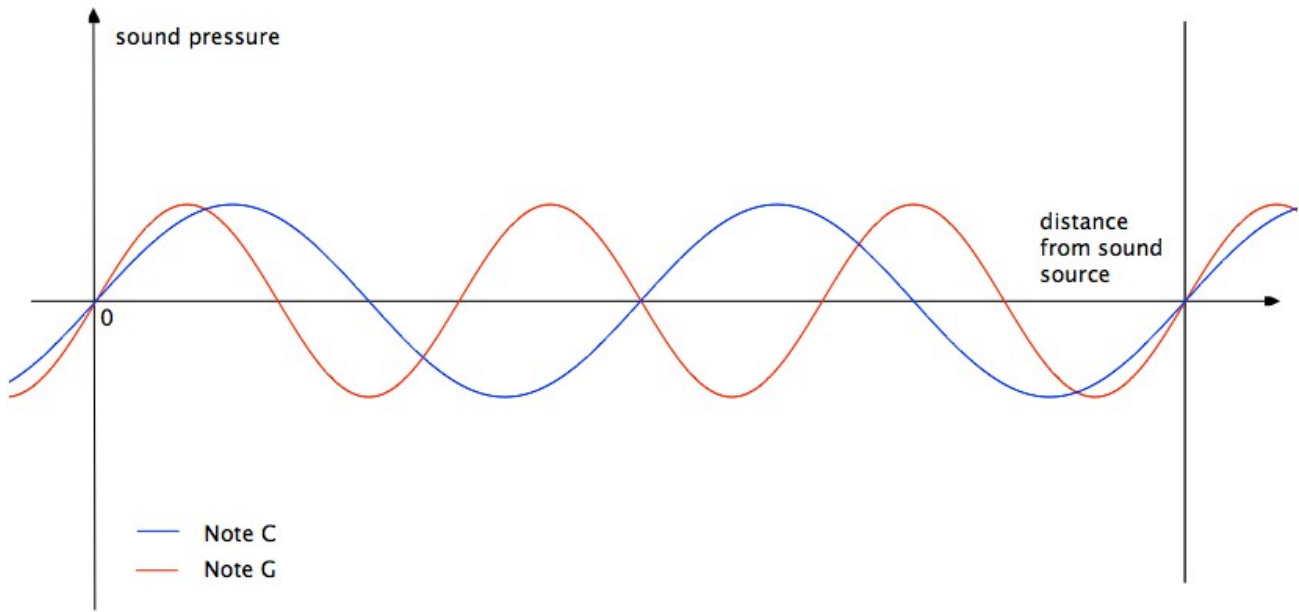

Figure 2: Graph of sound pressure versus distance from sound source in fixed time.

If we imagine the propagation of sound pressure in space at a fixed time (near the sound source), then we get the same figure (fig. 2).

The notes C and E (the big third) have lengths of sound waves differing by 5/4 times. Their graphs intersect on the abscissa axis when the sound pressure function of the note C makes 4 vibrations, and the E note function - 5. That is why the perfect fifth is more consonant than the big third.

Color, like sound, is also a wave. In the case of constructing a consonant interval for a color, we are not limited by a small set of notes, but are limited by the wavelength limits of the visible light, just as the sound is limited by the wavelength limits of the audible sound. The absence of restriction by a small set of notes leads to the fact that we can significantly expand the list of intervals - 3/2, 3/4, 2/5, 3/5, 4/5, ...

## 1. THE WAVE METHOD OF BUILDING COLOR PALETTE

Consider building of color palette for spectral and non-spectral colors.

The spectral color is a color having a certain wavelength. To build a color palette, let's take, first, the most consonant color to it - it's a color with a wavelength that differs in 1.5 times, but does not go beyond the visible spectrum. Further, similarly, we will take less consonant intervals until we reach the desired number of colors in the desired palette.

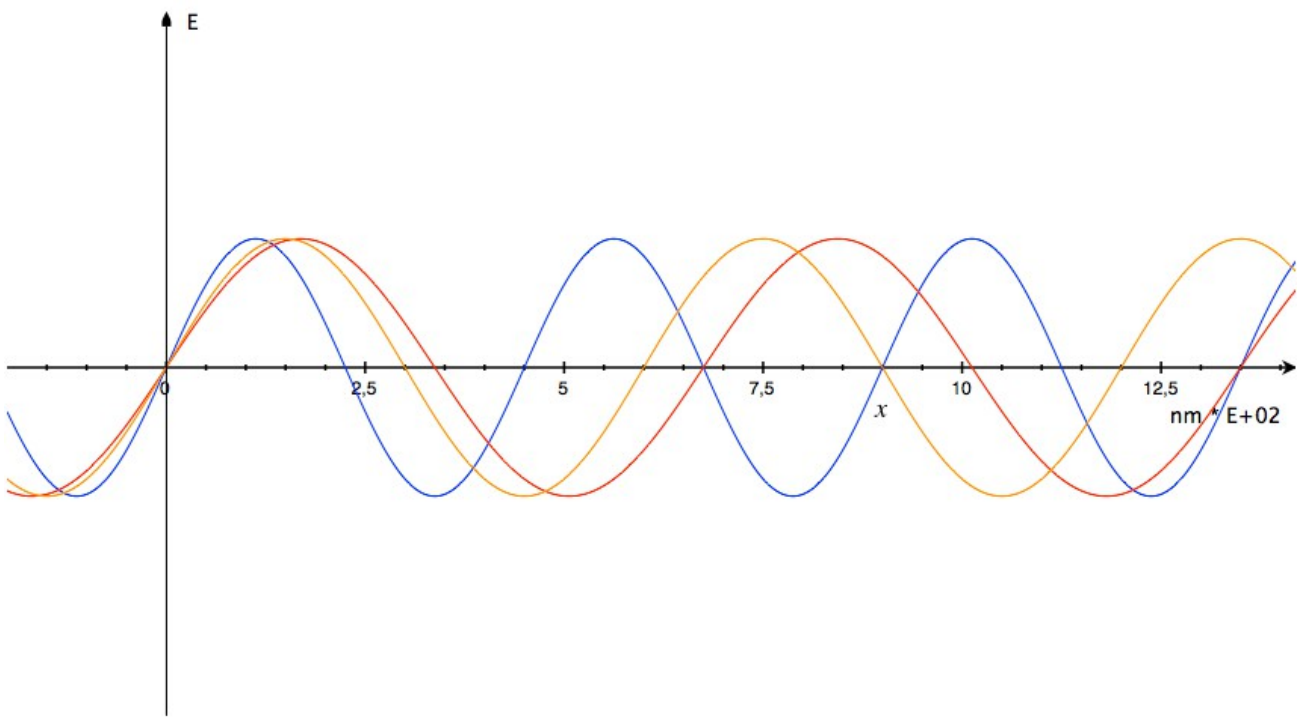

Figure 3: Graph of electric field intensity (E) versus distance from source in fixed time.

Take, for example, a blue color with a wavelength of 450 nm. The color whose wavelength is less than 1.5 times exceeds the scope of visible radiation. The color with a wavelength greater than 1.5 times (675 nm.) is a red color. The color with a wavelength larger by 3/4 times (600 nm.) is an orange color. As a result, we got the following color palette: the main color is blue, the most suitable color is red, a little less suitable for the blue color is orange (fig. 3). The same results can be obtained by operating frequencies instead of the wavelengths.

Also in music there is the concept of tune [5, p. 60]. The combination of notes can sound not only harmoniously, but also have a shade - a tune (ionian, dorian, phrygian, lydian, ...). Similar feelings can be transferred to the color palette by using the corresponding proportions when constructing it.

Nonspectral colors include colors that are not contained in the spectrum and consist of several spectral colors. Proceeding from the Grassmann additivity law [4, p. 25] it follows that in the case of selecting a color palette for non-spectral colors the same operations should be performed on its constituent colors preserving the proportions and considering the wavelength boundaries of the visible spectrum.

Let us consider a more recent phenomenon of consonance between two non-spectral colors. The musical sound consists of elementary tones, since along with the oscillation of the sound source itself as a whole, its parts also oscillate [5, p. 9]. Vibrations of the parts of the vibrating body give rise to weak prisms - overtones absorbed in the basic tone [5, p. 9]. The scale of simple tones of the corresponding amplitudes forming a complex sound is called the frequency spectrum [5, p. 9]. All elementary tones that enter into a complex sound are called harmonics [5, p. 10]. The degree of consonance of the interval is determined by the number of coinciding harmonics of the spectra of both notes: the greater number of coinciding harmonics - the more consonant interval [5, p. 51 - 52].

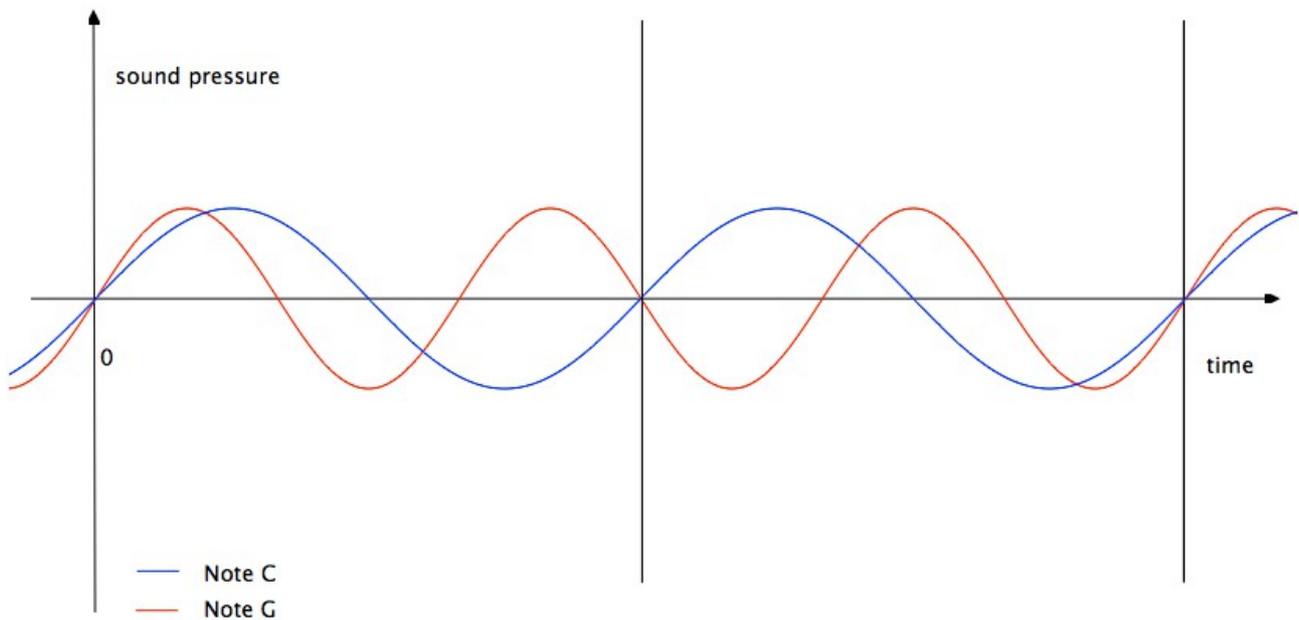

Figure 4: Graph of sound pressure versus time in fixed distance from sound source.

Proceeding from this, we believe that the essence of the phenomenon of harmony (consonance) is in the synchronous state of rest (energy is equal to zero) of both waves. On the graphs, this state of rest is displayed in the intersection of two wave functions on the abscissa (time) axis (fig. 4). And the degree of consonance of two wave functions is determined by the number of such intersections per unit of time (or length, under the condition of the same propagation speed): the more - the more consonant. This concept of degree of consonance extends also to non-spectral colors, since they are also wave functions.

## 2. APPLICATION OF THE WAVE METHOD IN COMPUTER GRAPHICS

A human is able to see colors with wavelengths in the range of 380 - 780 nm. [1, p. 30]. Any color feeling (red, yellow, green, brown, ...) can be obtained by mixing in a certain proportion of three independent colors (Grassmann's first law). The independence of colors, according to Grassmann, is that the color feeling caused by one of these three colors can not be obtained by mixing the other two colors in any proportions [6, p. 3]. It was noticed that it is most convenient to operate with red, green and blue color [6, p. 5]. Almost all modern monitors work on this principle.

In 1931 the International Lighting Congress (CIE) adopted a characteristic of the color properties of the average (standard) observer, based on the results obtained in 1926 — 1930 by Wright and Guild [4, p. 56]. The basis of this colorimetric standard is the following colors: 700 nm. (red), 546.1 nm (green) and 435.8 nm. (blue) [6, p. 7 - 8] (RGB system). The received characteristic contains the relationship between the resulting wavelength of the mixture and the amount of red, green, and blue colors in a given mixture [4, p. 62].

Later, for comfort calculations, the International Lighting Congress introduced the abstract system CIE XYZ [6, p. 16], based on unreal colors. This coordinate system is very comfort for the transition from one system to another [6, p. 17]. Also, the wavelengths of visible light and the corresponding coordinates of the CIE XYZ mixture were calculated [4, p. 75, 78], based on the results obtained for the RGB system.

To reproduce the same color feelings on different output devices (monitor or printer), each such device has its own color profile, which contains its relationship with the abstract CIE XYZ system [7, p. 105 - 107]. In other words, the color profile is used for the possibility of switching between different color systems (sRGB, AdobeRGB, ...). The most common color space is the sRGB system. Next, consider the implementation of the wave method for this system.

Find the wavelengths of red, green and blue in the sRGB system. To do this we use the transition formulas from the sRGB system to CIE XYZ [8, 9]. As a result, we get the following coordinates in the CIE XYZ system: red - (0.412456, 0.212673, 0.019334), green - (0.357576, 0.715152, 0.119192), blue - (0.180437, 0.072175, 0.950304).

Using tables containing the wavelengths of visible light and the corresponding coordinates of the CIE XYZ mixture [4, p. 75, 78], we find the wavelengths for red, green and blue colors in the sRGB system: red - 611.4 nm, green - 549.1 nm, blue - 464.2 nm.

Now let's find the most consonant color for each of the above colors (within the visible radiation range of 380 - 780 nm.). For red - this is 611.4 / 1.5 = 407.6 nm. For green multiplication or division by 1.5 (by 3/2) goes beyond visible radiation. For this take 3/4: 549.1 * 3/4 = 411.825 nm. or 549.1 / 3/4 ≈ 732.133 nm. For blue - this is 464.2 * 1.5 = 696.3 nm.

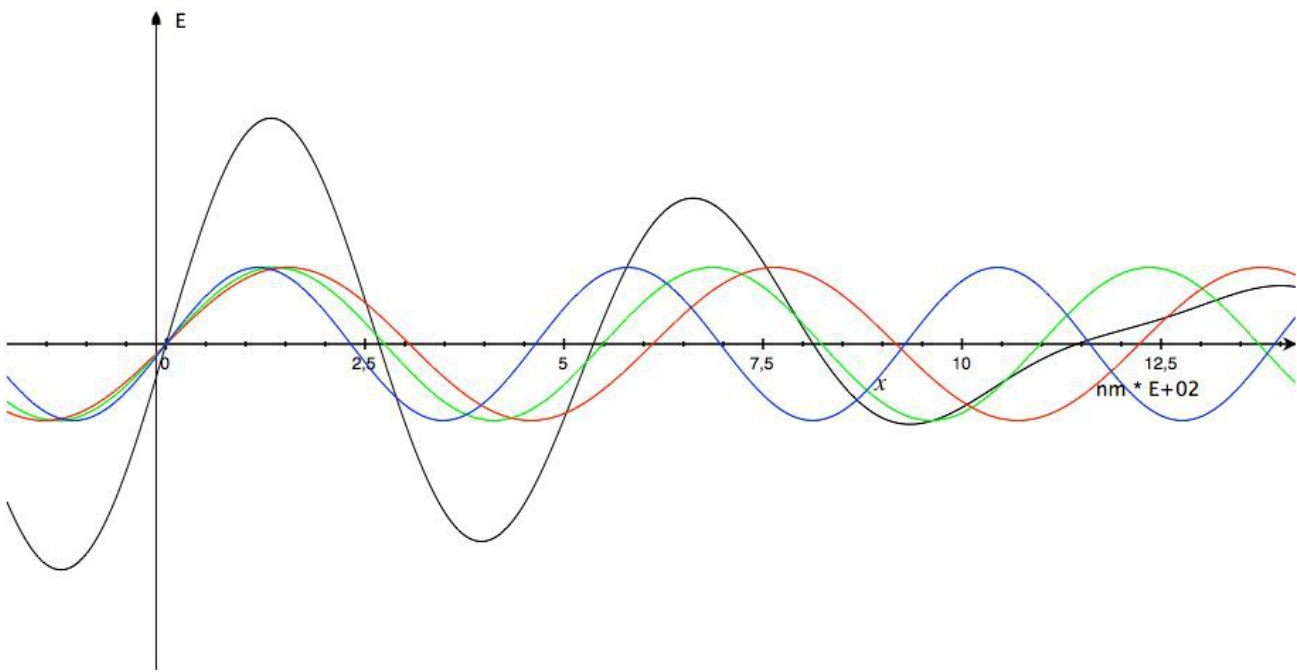

Figure 5: Graph of electric field intensity (E) versus distance from source in fixed time.

Suppose that the initial mixture consists of equal parts of red, blue and green (fig. 5). The corresponding coordinates in the sRGB system are (1, 1, 1). The resulting curve will have the next equation:

$$y = \sin(x * 2\pi / 6.114) + \sin(x * 2\pi / 5.491) + \sin(x * 2\pi / 4.642). \qquad (1)$$

The resulting curves in the consonant mixture:

$$y = \sin(x * 2\pi / 4.076) + \sin(x * 2\pi / 4.118) + \sin(x * 2\pi / 6.963), \qquad (2)$$

$$y = \sin(x * 2\pi / 4.076) + \sin(x * 2\pi / 7.321) + \sin(x * 2\pi / 6.963). \qquad (3)$$

Also consider the consonant mixture in which the green color remained unchanged:

$$y = \sin(x * 2\pi / 4.076) + \sin(x * 2\pi / 5.491) + \sin(x * 2\pi / 6.963). \qquad (4)$$

Analyzing the intersections of the functions (2) - (4) with the function (1) on the abscissa axis, and also near the abscissa axis, from 0 to 10,000 nm. we decided on option (3), although (2) and (4) also fit in the role of consonant mixtures.

Let us find the corresponding CIE XYZ coordinates for the wavelengths involved in equation (3): 407.6 nm. - (0.1728, 0.0048, 0.8224), 732.1 - (0.0013, 0.0005, 0.0000), 696.3 nm. - (0.7347, 0.2654, 0.0000). The corresponding coordinates in the sRGB system are 407.6 nm. - (0.412554, -0.387623, 0.942622), 732.1 nm. - (0.075942, -0.025863, -0.008750), 696.3 nm. - (1.361850, -0.496422, -0.140190). Thus, the formula for obtaining the consonant color (r1, g1, b1) for the color in the sRGB system (r, g, b) is follows:

r1 =   0.412554 * r +  0.075942 * g +  1.361850 * b,

g1 = - 0.387623 * r -  0.025863 * g -  0.496422 * b,

b1 =   0.942622 * r -  0.008750 * g -  0.140190 * b.

In the event that any of the parameters exceeds one - we proportionally reduce all the values so that this parameter becomes equal to one. If any of the parameters is negative - we equate it to zero.

Now, similarly, we find the following consonant color for each of the above colors (within the visible radiation range of 380 - 780 nm.). For red - this is 611.4 * 3/4 = 458.55 nm. For green - this is 549.1 * 3/4 = 411.825 nm. or 549.1 / 3/4 ≈ 732.133 nm. For blue - this is 464.2 * 4/3 = 618.933 nm. Choosing a consonant color for green color with a wavelength of 732.133 nm., we obtain the following results: 458.55 nm. - (0.153969, -0.265282, 0.941624), 732.133 nm. - (0.075942, -0.025863, -0.008750), 618.933 nm. - (1.291906, -0.327664, -0.186152). As a result we get:

$r2 = 0.153969 * r + 0.075942 * g + 1.291906 * b,$

$g2 = - 0.265282 * r - 0.025863 * g - 0.32766 * b,$

$b2 = 0.941624 * r - 0.008750 * g - 0.186152 * b.$

Similarly, the next consonant colors can be obtained for the ratios 2/5, 3/5, 4/5, ...

This method can find wide application in various branches of design. For example, in web-design, provided the color profile for the site is specified in css-markup (color-profile: sRGB). The results of this work are on the site wavepalette.com [10] and will be regularly updated and, if necessary, adjusted.

## CONCLUSIONS

In this article we have described and justified from the physical point of view the wave method of constructing the color palette developed by us. Was described our understanding of the essence of the phenomenon of harmony. Also in the article was described the algorithm for implementing the wave method for the sRGB color model in computer graphics. This method can be widely used in design.